\documentclass[letterpaper]{article} 
\usepackage{aaai25}  
\usepackage{times}  
\usepackage{helvet}  
\usepackage{courier}  
\usepackage[hyphens]{url}  
\usepackage{graphicx} 
\usepackage{natbib}  
\usepackage{caption} 
\usepackage{algorithm}
\usepackage{algorithmic}
\usepackage{amsmath}
\usepackage{amssymb}
\usepackage{multirow}
\usepackage{subcaption}

\usepackage{newfloat}
\usepackage{listings}
\DeclareCaptionStyle{ruled}{labelfont=normalfont,labelsep=colon,strut=off} 
\floatstyle{ruled}
\newfloat{listing}{tb}{lst}{}
\floatname{listing}{Listing}
\nocopyright
\title{DREAM: A Dual Representation Learning Model for Multimodal Recommendation}
\author{
    Kangning Zhang\textsuperscript{\rm 1},
    Yingjie Qin\textsuperscript{\rm 2},
    Jiarui Jin\textsuperscript{\rm 2},
    Yifan Liu\textsuperscript{\rm 1},
    Ruilong Su\textsuperscript{\rm 2},
    Weinan Zhang\textsuperscript{\rm 1},
    Yong Yu\textsuperscript{\rm 1}
}
\affiliations{
    \textsuperscript{\rm 1}Shanghai Jiao Tong University\\
    \textsuperscript{\rm 2}Xiaohongshu,China\\
    zhangkangning@sjtu.edu.cn, huanling@xiaohongshu.com
}

\usepackage{bibentry}
\begin{document}

\maketitle

\begin{abstract}
Multimodal recommendation focuses primarily on effectively exploiting both behavioral and multimodal information for the recommendation task. However, most existing models suffer from the following issues when fusing information from two different domains:
(1) Previous works do not pay attention to the sufficient utilization of modal information by only using direct concatenation, addition, or simple linear layers for modal information extraction.
(2) Previous works treat modal features as learnable embeddings, which causes the modal embeddings to gradually deviate from the original modal features during learning. We refer to this issue as Modal Information Forgetting. 
(3) Previous approaches fail to account for the significant differences in the distribution between behavior and modality, leading to the issue of representation misalignment. To address these challenges, this paper proposes a novel \textbf{D}ual \textbf{RE}present\textbf{A}tion learning model for \textbf{M}ultimodal Recommendation called \textbf{DREAM}.
For sufficient information extraction, we introduce separate dual lines, including Behavior Line and Modal Line, in which the Modal-specific Encoder is applied to empower modal representations. To address the issue of Modal Information Forgetting, we introduce the Similarity Supervised Signal to constrain the modal representations. Additionally, we design a Behavior-Modal Alignment module to fuse the dual representations through Intra-Alignment and Inter-Alignment.
Extensive experiments on three public datasets demonstrate that the proposed DREAM method achieves state-of-the-art (SOTA) results. The source code will be available upon acceptance.

\end{abstract}

%

\section{Introduction}
Personalized recommendation has become prevalent in multimedia content-sharing platforms to help users discover items of interest. 
To alleviate the data-sparsity problem, multimodal information has been introduced into the recommendation system. 
We claim that the main challenge in current multimodal recommendation models is How to effectively exploit the behavior information from interaction data and multimodal information from multimedia data?


\begin{figure}[h]
\setlength{\belowcaptionskip}{-0.5cm}
  \centering
  \includegraphics[width=0.50\textwidth]{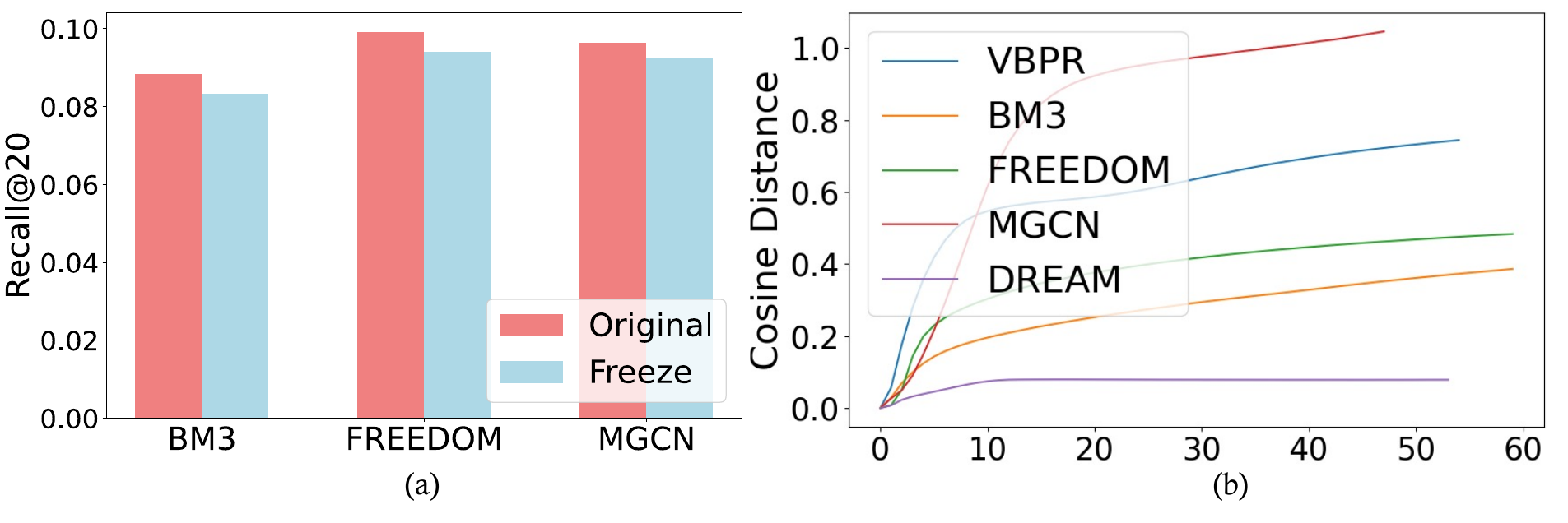}
  \caption{
  (a) When Modal Embeddings are frozen, the model performance drops;
  (b) The Cosine Distance between learnable modal embeddings and original modal features. Compared to previous works (e.g. VBPR, BM3, FREEDOM, MGCN), the modality embeddings in DREAM converge more quickly and effectively maintain the original modality information through Similarity Supervised Signal.}
  \label{fig:information_shift}
\end{figure}


Though making significant progress, previous multimodal recommendation models suffer from three issues. 
The first issue is the \textbf{unsufficient utilization of modal information}. Many studies~\cite{wei2019mmgcn,GRCN,BM3} have adopted various neural networks to strengthen the capture of behavioral information and improve the learning of ID representation. Although some work has utilized multimodal information to construct item-item relation graphs~\cite{lattice,Freedom}, they still serve to learn the representation of ID. For modal information, many works~\cite{vbpr,BM3,Freedom} try to directly concatenate modal features with behavioral representations or use simple linear layers to map modal features to the behavioral domain. When the behavioral representation sufficiently captures behavioral signals, the modal information embodied in the modal representation is inadequate.

Secondly, previous works~\cite{BM3,Freedom,MGCN} have treated modal features as learnable embeddings, which poses the risk of progressive attenuation of modal information during the learning process. In this paper, we refer to this issue as \textbf{Modal Information Forgetting}. However, simply freezing the modal embeddings cannot effectively solve this problem, as shown in Figure \ref{fig:information_shift}(a), where the frozen modal embeddings lead to a drop in model performance.
We use the cosine distance between the learnable modal embeddings and the original modal features as a proxy metric to quantify the loss of modal information. As shown in Figure \ref{fig:information_shift}(b), the mainstream models exhibit a gradual divergence of the learned modal embeddings from the original multimodal features over the course of the training process. In contrast, DREAM demonstrates rapid convergence of the modal embeddings towards the initial modal features, effectively retaining the original modal information throughout the learning procedure.

\begin{figure}[h]
\setlength{\belowcaptionskip}{-0.3cm}
  \centering
  \includegraphics[width=0.48\textwidth]{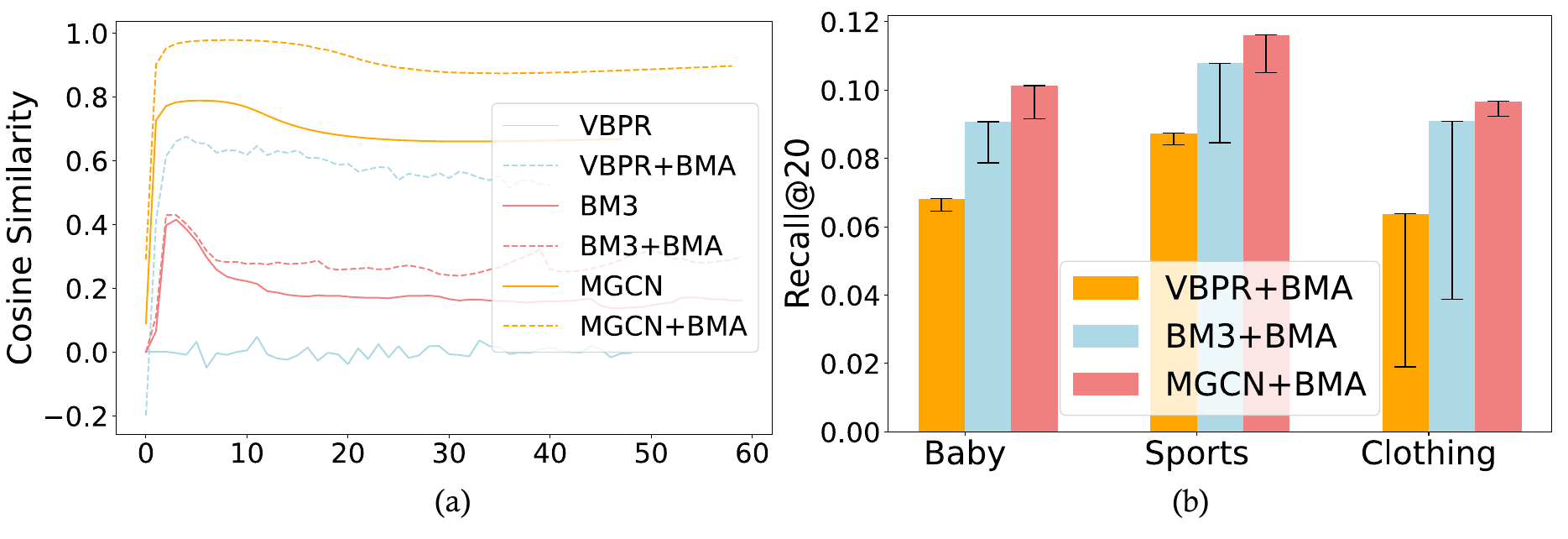}
  \caption{
  (a) The cosine similarity between dual domains of previous works.
  (b) The performance improvement of previous works through introducing the BMA, with model structure and hyperparameters unchanged.
  }
  \label{fig:bma_introduction}
\end{figure}

The \textbf{lack of alignment between behavior and modality representations} is another significant issue. 
Previous work has attempted various methods to blend their information: 
concatenating~\cite{vbpr}, utilizing graph neural networks~\cite{wei2019mmgcn,GRCN,dualgnn}, introducing self-supervised methods~\cite{BM3, SLMRec}. However, we claim that if representations from dual domains cannot be properly aligned before integrating, the learning capability of the model might be reduced significantly. Figure \ref{fig:bma_introduction}(a) shows the changes in cosine distance of behavior-related and modal-related representations during the learning process for various models. These models still exhibit significant distributional differences between dual domains. Therefore, we introduced our Behavior-Modal Alignment module (to be discussed in detail later) based on these models. Figure \ref{fig:bma_introduction}(a) shows that BMA effectively increases the cosine similarity between dual representations. Additionally, Figure \ref{fig:bma_introduction}(b) demonstrates the performance improvements brought by aligning the behavior-modal representation for these models. 


In this paper, 
we introduce DREAM, a Dual representation learning model for Multimodal recommendation, where one representation only learns from the interaction data (behavior information) and another representation fully focuses on the pre-extracted multimodal features. In our model, there are two parallel lines to calculate dual representations. \textit{On the Behavior Line}, we perform LightGCN~\cite{he2020lightgcn} on the normalized user-item interaction graph to generate Behavior representations for users and items. \textit{On the Modal Line}, we design the Modal-specific Encoder for fine-grained representation extraction, which includes Modal-specific filter gates and relation graphs.
Previous works~\cite{Freedom,MGCN} only consider modal representation learning at the item level. In DREAM, we attempt to extend this to the user level, innovatively constructing user-level Modal-specific filter gates and Modal-specific U-U graphs. The Modal-specific filter gates denoise the preference-irrelevant modality noise contained in original multimodal features. Meanwhile, the Modal-specific relation graphs connect different instances from the perspective of semantic similarity. Through these two key components, we are able to model the modal representations of both users and items in a more fine-grained manner, further enhancing the capacity to utilize multimodal information.

To mitigate the problem of Modal Information Forgetting illustrated in Figure \ref{fig:information_shift}, we introduce the Similarity Supervised Signal (S3). The S3 constrains the Modal Representation output by the Modal Line to maintain the similarity information embodied in the original modality features. Through the introduction of the S3, the modal embedding is constrained during the learning process. As a result, the modal embedding converges quickly on the basis of the initial modal features, and does not experience significant deviation, as shown in Figure \ref{fig:information_shift}(b).



To facilitate alignment between behavior representation and modality representation, we designed the Behavior-Modal Alignment (BMA) module, which utilizes contrastive loss for Intra-Alignment and Inter-Alignment. The Intra-Alignment refers to the alignment of both user and item behavioral representations, as well as the alignment of user and item modality representations. It aims to ensure that the representations within each domain (behavioral and modality) are aligned properly. Inter-Alignment, in contrast to Intra-Alignment, focuses on aligning the behavioral and modality representations of users and items separately. It aims to ensure that the behavior representation of a user (item) is aligned with the corresponding modal representation of the same user (item). Through our alignment process, the behavior and modal representations are brought into the same latent space. Furthermore, Behavior-Modal Alignment can be easily incorporated into other multimodal recommendation models to further enhance their recommendation performance, as shown in Fig. \ref{fig:bma_introduction}(b). Finally, we only need to perform a simple summation of the Behavior and Modal representations to merge information and complete recommendation task.
We conduct extensive experiments on three public datasets and DREAM achieves state-of-the-art (SOTA) performance, which demonstrates its effectiveness for multimodal recommendation.



We summarize our main contributions as follows:
\begin{itemize}
\item We propose a dual representation learning model called DREAM, which includes dual lines and symmetrically computes the Behavior and Modal representations. Especially, we introduce Modal-specific Encoder including filter gates and relation graphs to empower the learning of modal representations for both users and items. 
\item To our best knowledge, we are the first to introduce the problem of Modal Information Forgetting, and propose the Similarity Supervised Signal to guide the modal representations in maintaining the initial modality similarity information during the learning process. This approach helps to mitigate the issue of information loss, while also improving the overall recommendation performance.
\item We introduce the Behavior-Modal Alignment (BMA) module to tackle misalignment problems through both Intra-Alignment and Inter-Alignment. Furthermore, the BMA module can be seamlessly integrated into other multimodal recommendation models, enhancing their recommendation performance.
\end{itemize}

\section{Related Works}
\subsection{Multi-modal Recommendation}
Most early multimodal recommendation models utilize deep learning to explore user preferences. VBPR~\cite{vbpr} obtain item representation by concatenating the latent visual features and id embedding. Deepstyle~\cite{liu2017deepstyle}
augments the representations with both visual and style features. VECF~\cite{chen2017attentive} utilizes the VGG model~\cite{VGG} to capture the user attention on different image regions. Recently, there has been a line of works that introduce GNNs into multimodal Recommendation~\cite{wu2022graph}. MMGCN~\cite{wei2019mmgcn} construct modality-specific user-item bipartite graph to capture users' preference for specific modality. DualGNN~\cite{dualgnn} introduces a user co-occurrence graph. GRCN~\cite{GRCN} scores the affinity between users and items and identifies the false-positive edges. LATTICE~\cite{lattice} create the item-item graph for each modality and fuse them together to obtain a latent item graph. FREEDOM\cite{Freedom} further freezes the item-item graph for effective and efficient recommendation. 


\subsection{Compared with Previous works}
There has been substantial prior works~\cite{CLIP,li2022blip,align_before_fusion,beit,li2023blip2} exploring the alignment of multimodal information, and the following studies have also attempted to extend these approaches to the domain of multimodal recommendation.
SLMRec~\cite{SLMRec} introduces data augmentation upon multi-modal contents to generate multiple views of individual items. To align different modalities, SLMRec introduce Modal-Agnostic task and Modal-Specific task in multi-task framework. 
BM3~\cite{BM3} utilizes the simple dropout technique to generate contrastive views of multimodal features and uses three contrastive loss functions to optimize representations. BM3 utilize cosine similarity as supervise signal and try to align modal features with id features.   
AlignRec~\cite{AlignRec} decomposes the recommendation objective as three alignments, where each alignment is characterized by a specific objective function. Specifically, AlignRec emphasizes the alignment between various multimodal features, and extracts unified multimodal features through efficient pre-training.

Compared with these works, DREAM still has the following novelties: i) The alignment in DREAM is fully focused on dual domain between behavior and modality, which has been rarely studied in depth. About different modalities input, it utilizes Modal-specific Encoders to extract fine-grained representations for each modality; ii) We conduct a deeper investigation into the impact of aligning behavior and modality in previous works, and find that the Behavior-Modal Alignment (BMA) can be seamlessly integrated into multiple models, consistently leading to performance improvements; iii) We consider the problem of Modal Information Forgetting and propose Similarity Supervised Signal for this issue; IV) DREAM achieves optimal performance compared with previous alignment works.

\begin{figure*}[ht]
  \centering
  \includegraphics[width=\textwidth]{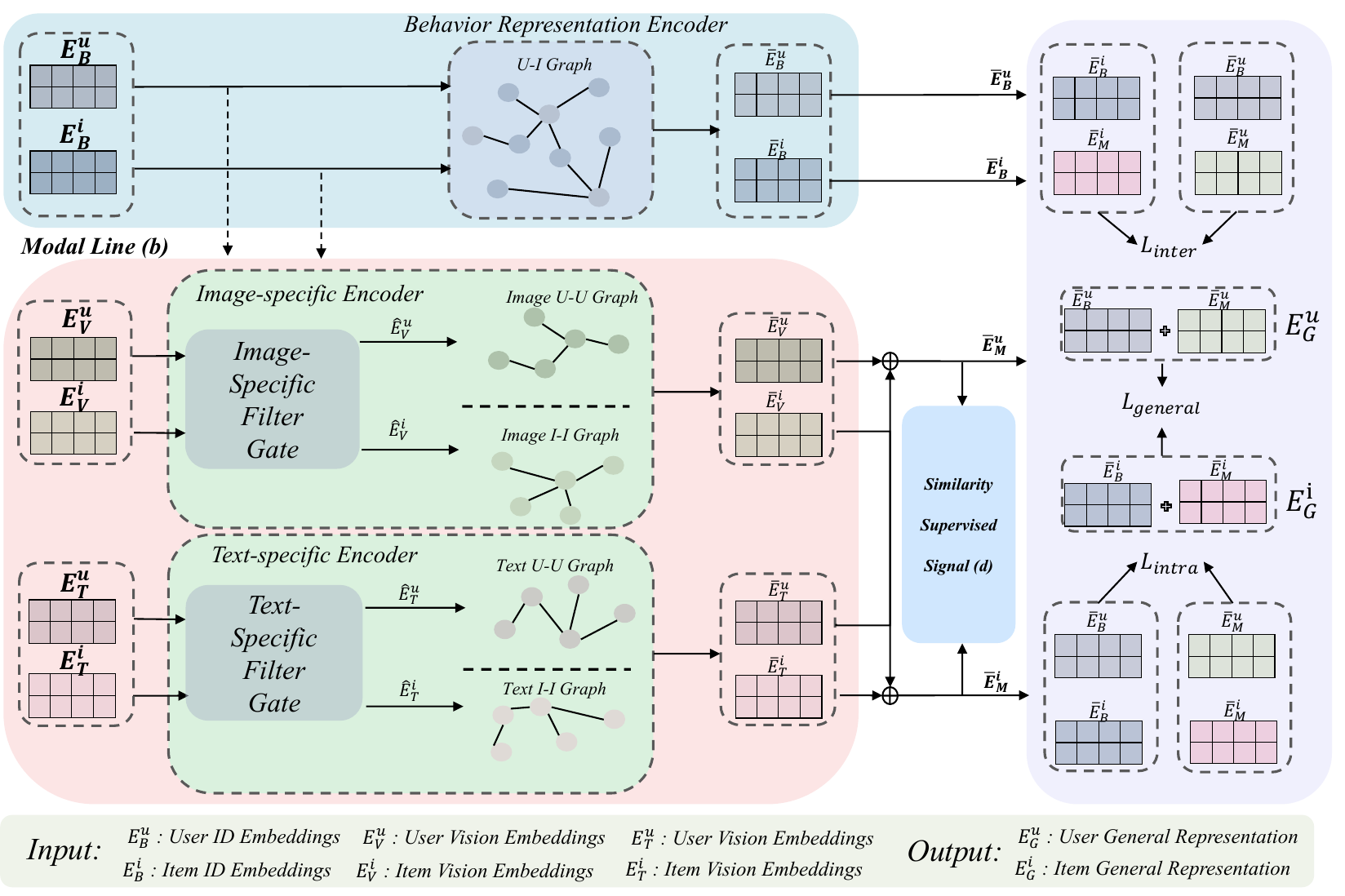}
   
  \caption{
  The overview of DREAM: (a) Behavior Line utilizes ID embedding for behavior representation learning. (b) Modal Line focuses on utilizing the multimodal features through Modal-specific filter gates (Fig. \ref{filter_gate}) and relation graphs. (c) Similarity-Supervised Signal (Fig. \ref{S3 figure}) constraints the learning of modal representations to mitigate the problem of Modal Information Forgetting. (d) Behavior-Modal Alignment module consists of Intra-Alignment and Inter-Alignment for representation alignment and information fusion.}
  \label{fig:structure}
\end{figure*}

\section{DREAM}

\subsection{Problem Definition}
Given a set of $M$ users $\mathcal{U} = \{ u_i\}_{i=1}^{M}$, a set of $N$ items $\mathcal{I} = \{ i_t\}_{t=1}^{N}$. Each item is associated with multimodal features $e_m^i \in \mathbb{R}^{d_m}$, where $m \in \{V, T\}$, $V$ and $T$ represent visual and textual modality respectively. 
User historical behaviour data is denoted as $R \in \mathbb{R}^{M \times N}$,  where $R_{ui}=1$ means user $u$ has interacted with item $i$, otherwise $R_{ui}=0$. Naturally, $R$ can be regarded as a sparse behavior graph $\mathcal{G}=(\mathcal{V}, \mathcal{E})$, where $\mathcal{V} =  \mathcal{U} \cup \mathcal{I}$ denotes the set of nodes and $\mathcal{E} = \{ (u,i) | u \in \mathcal{U}, i \in \mathcal{I}, R_{ui} =1\}$ denotes the set of edges. We use $\mathcal{I}_u=\{i| (u,i) \in \mathcal{E}\}$ to denote the set of items which a user $u$ has interacted with.
The purpose of the multimodal recommendation is to accurately predict users' preferences by ranking items for each user according to predicted preferences scores $y_{ui} = f_{\theta}(e_B^u, e_B^i, e_m^i, e_m^u)$, where $\theta$ means model parameters, $e_B^u$, $e_B^i$ mean ID embedding and $e_m^i$, $e_m^u$ mean modal features for user and item. In this work, the user modal feature $e_m^u$ can be easily calculated by $e_m^u=\frac{1}{|\mathcal{I}_u|} \sum_{i \in \mathcal{I}_u} e_m^i$.
\footnote{Specifically, we use $e$ to represent the embedding of an individual instance, and $E$ represent the concatenated embedding table. Superscripts $u$ and $i$ represent users or items, while subscripts $B$ and $m \in \{V,T\}$ represent behavior or multimodal domains.}


\subsection{Behavior-Line Representation Learning}\label{Behavior Line}
\subsubsection{\textbf{Behavior Representation Encoder}}
In our work, we use the Behavior representation to fully capture the behavior signal (e.g. click or purchase feedback). 
We construct a symmetric adjacency matrix $A \in \mathbb{R}^{|\mathcal{V}| \times |\mathcal{V}|}$ from the user-item interaction matrix $R$:
\begin{equation}
    A = \begin{pmatrix}
    0 & R\\
    R^T &0
    \end{pmatrix}
\end{equation} and each entry $A_{ui}$ of $A$ is set to 1 if user $u$ has interacted with item $i$, otherwise 0. The diagonal degree matrix of $A$ is denoted by $D$ and $D_{ii}=\sum_j A_{ij}$. 
We use $E_B^l \in \mathbb{R}^{|\mathcal{V}| \times d}$ to denote the behavior embedding at the $l$-th layer by stacking all the embedding of users and items at layer $l$. $E_B^0$ is the concatenation of $E_B^u$ and $E_B^i$. We use simplified graph convolutional layers in LightGCN~\cite{he2020lightgcn} to perform forward propagation:
\begin{equation}
    E_B^{l+1} = (D^{-\frac{1}{2}}AD^{-\frac{1}{2}})E_B^l
\end{equation}


We use mean function to aggregate all representations in hidden layers for Behavior user and item representation:
\begin{equation} \label{Behavior Line Equation}
    \begin{aligned}
        \overline{E}_{B} &= \textbf{Mean}(E_B^{0}, E_B^{1}, ..., E_B^{L}) 
    \end{aligned}
\end{equation}
Finally, $\overline{E}_{B}^u$ and $\overline{E}_{B}^i$ can be obtained by respectively taking the first $M$ rows and the last $N$ rows of $\overline{E}_{B}$.


\subsection{Modal-Line Representation Learning}\label{Modal Line}
\subsubsection{\textbf{Modal-specific Encoder}}

\begin{figure}[h]
\setlength{\belowcaptionskip}{-0.5cm}
  \centering
  \includegraphics[width=0.48\textwidth]{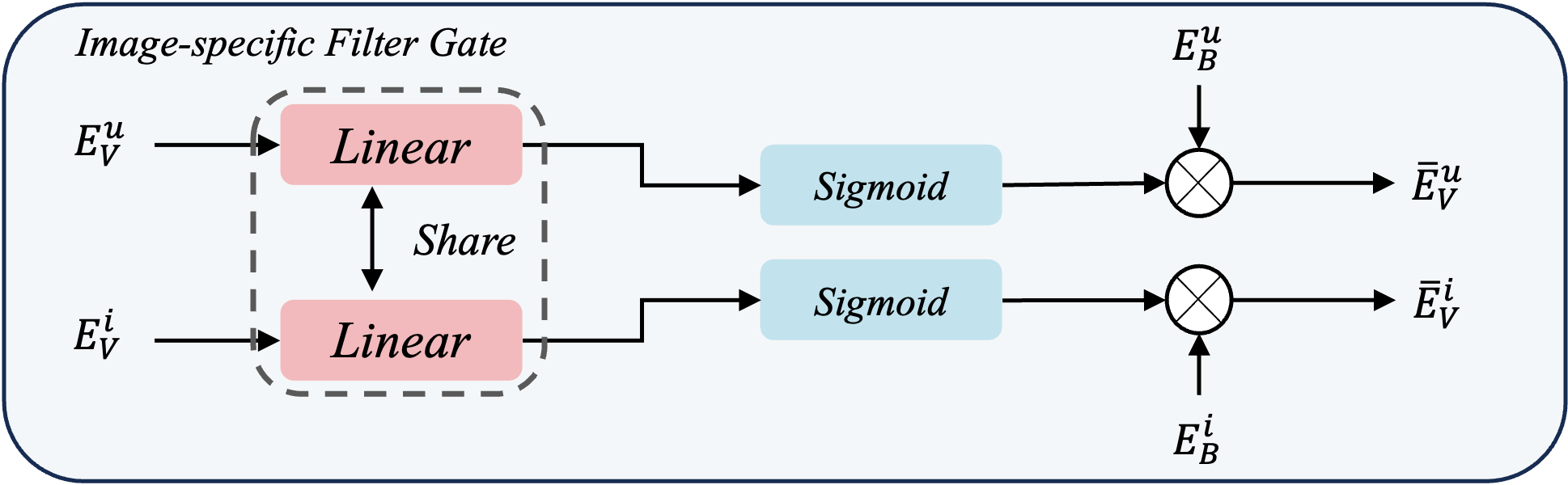}
  \caption{
  The structure of Image-specific Filter Gate in Vision modality. The Text-specific Filter Gate is similar. 
  }
  \label{filter_gate}
\end{figure}


Let $E_V^i \in \mathbb{R}^{N \times d_v}$, $E_T^i \in \mathbb{R}^{N \times d_t}$, $E_V^u \in \mathbb{R}^{M \times d_v}$, and $E_T^u \in \mathbb{R}^{M \times d_t}$ denote the item raw feature tables in vision and text modalities, and the user modal feature tables, respectively.
The Modal-specific Encoder includes the Modal-specific Filter Gate and Modal-specific relation graphs. We herein only illustrate the process for the vision modality, and text modality is similar. Inspired by MGCN~\cite{MGCN}, we design the modal-specific Filter Gate for denoising the initial multimodal features, which is showed in Figure ~\ref{filter_gate}. The Vision-specific Filter Gate takes $E_V^i$, $E_V^u$ and $E_B^i$, $E_B^u$ as inputs and output the filtered vision representation $\hat{E}_V^i$ and $\hat{E}_V^u$:
\begin{align}
    \begin{aligned}
        \hat{E}_V^i &= \text{Gate}_v(E_V^i, E_B^i) = \sigma(E_V^i W_2 + b_2) \odot E_B^i \\
        \hat{E}_V^u &= \text{Gate}_v(E_V^u, E_B^u) =
        \sigma(E_V^u W_2 + b_2) \odot E_B^u
    \end{aligned}
\end{align} where $W_2 \in \mathbb{R}^{d_v \times d}$ and $b_2 \in \mathbb{R}^d$ are learnable parameters, $\odot$ denotes the element-wise product and $\sigma$ is the sigmoid function. According to MGCN~\cite{MGCN}, the Filter Gate help to purify the preference-irrelevant modality noise contained in multimodal information. Based on this, we extend filter gate to user level through sharing the linear layer within each modality.

Previous works~\cite{Freedom,MGCN} have focused on exploring the modal item-item graphs, but they neglect the impact of the modal user-user graphs. Therefore, we propose modal-specific relation graphs that contain not only item-item graphs but user-user graphs, which help to better capture modal semantic information. Specifically, we first calculate the similarity score between item or user pair on their raw modal features. To make the relation graph sparse, we only keep the top-$k$ similar items (users) of every item (user) and convert the weighted graph into an unweighted graph. Finally, we normalize the discretized graph. The calculation procedure about item-item graph in vision modality is as follows:
\begin{equation}\label{Modal graph}
    \begin{aligned}
     S_V^{ij} &=  \frac{(e_V^i)^T e_V^j}{\|{e_V^i}\|
     \|e_V^j\|} \\
     \hat{S}_V^{ij} &=\begin{cases} 
    1
    &S_V^{ij} \in \text{topk}(S_V^i) \\
    0 
    &\text{otherwise} 
    \end{cases} \\ 
    \tilde{S}_V &= (D^v)^{-\frac{1}{2}}\hat{S}_V(D^v)^{-\frac{1}{2}}
    \end{aligned}
\end{equation} $\tilde{S}_V$ is the final item-item relation graph in modality $V$. We use a similar way to construct user-user relation graph $\tilde{J}_V$ in modality $V$. The graph convolution over the item-item relation graph and user-user relation graph in modality $V$ is:
\begin{equation}
\begin{aligned}
    \overline{E}_V^i = \tilde{S}_V \hat{E}_V^i
    \quad
    \overline{E}_V^u =  \tilde{J}_V \hat{E}_V^u
\end{aligned}
\end{equation}



Finally, we fuse vision and text modality representation with modality-specific weight to obtain Modal Representation:
\begin{equation} \label{Modal Line Equation}
\begin{aligned}
    \overline{E}_{M}^u &= \alpha_V \overline{E}_V^u + \alpha_T \overline{E}_T^u \\
    \overline{E}_{M}^i &= \alpha_V \overline{E}_V^i + \alpha_T \overline{E}_T^i
\end{aligned}
\end{equation}
where  $\alpha_V \in [0, 1]$ and $\alpha_T = 1-\alpha_V$.

\begin{figure}[h]
  \centering
  \includegraphics[width=0.48\textwidth]{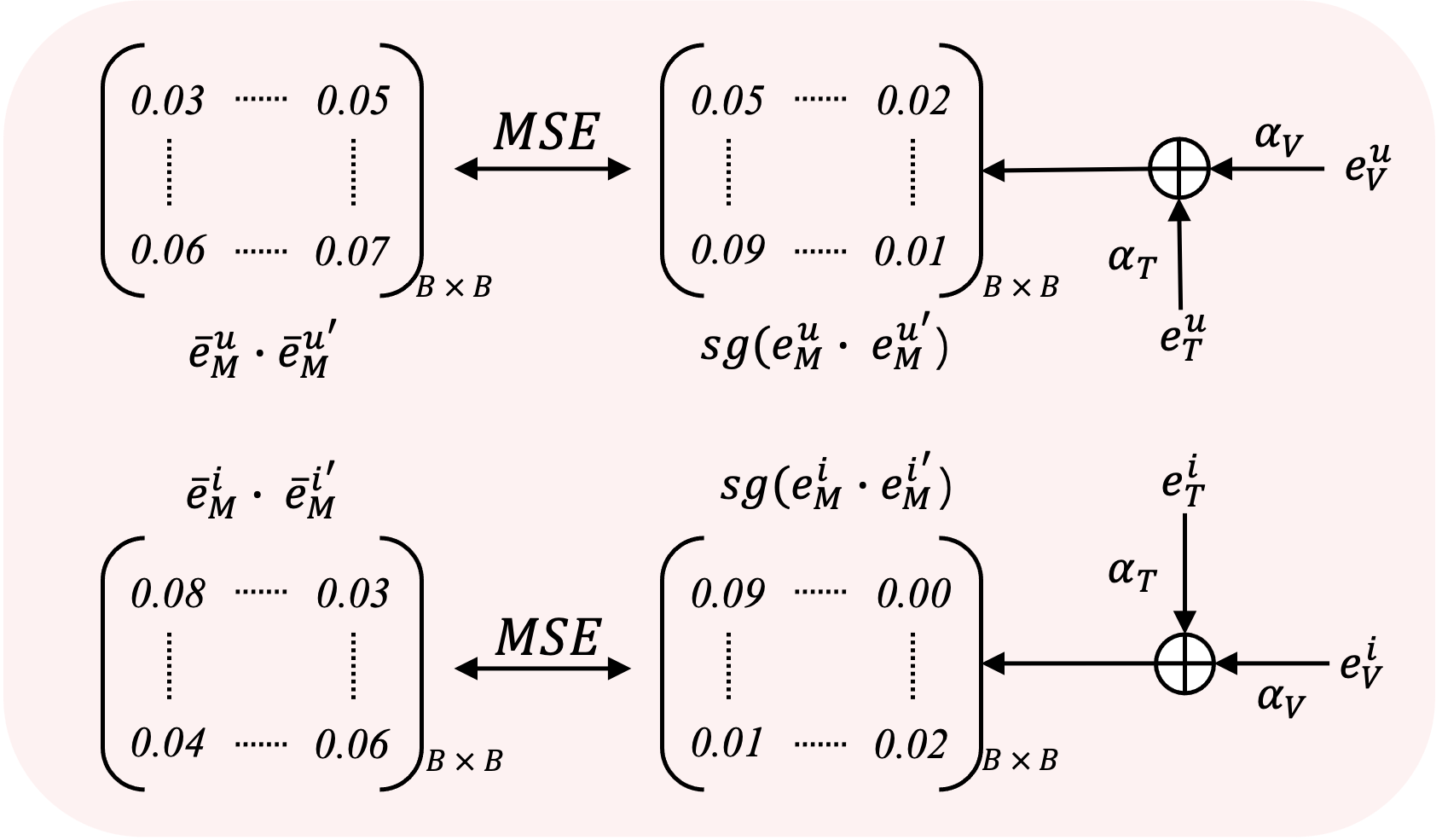}
  \caption{
  The Similarity Supervised Signal (S3) ensure that the batch-scale similarity matrix derived from modal representation is as similar as possible to the similarity matrix computed from the original modal features through Mean Square Error (MSE). 
  }
  \label{S3 figure}
\end{figure}
\subsubsection{\textbf{Similarity Supervised Signal}} \label{S3 section}
Previous works suffer from the problem of Modal Information Forgetting, with results shown in Fig. \ref{fig:information_shift}.
We introduce the Similarity-Supervised Signal (S3) to help the Modal representations retain the similarity information that exists in their original modal features:
\begin{equation}
    \begin{aligned}
    e_M^i &= \alpha_V e_V^i + \alpha_T e_T^i \\
    e_M^u &= \alpha_V e_V^u + \alpha_T e_T^u \\
    \mathcal{L}_{\text{S3}} &= \|\overline{e}_M^i \cdot \overline{e}_{M}^{i'} - sg(e_M^i \cdot e_M^{i'})\|_{2} \\
    &+ \|\overline{e}_M^u \cdot \overline{e}_{M}^{u'} - sg(e_M^u \cdot e_M^{u'})\|_{2}
    \end{aligned}
\end{equation}
where $\overline{e}_M^i$,$\overline{e}_M^u$ is from Eq. (\ref{Modal Line Equation}). $(i,i')$ and $(u,u')$ are item/user pair from the same batch, and $sg(\cdot)$ means stop gradients. In the concrete implementation, we compute the loss $\mathcal{L}_{S3}$ at the batch matrix level, as illustrated in Figure \ref{S3 figure}.

\subsection{Behavior-Modal Alignment} \label{BMA section}

\subsubsection{\textbf{Intra-Alignment}}
The Intra-Alignment includes Behavior Intra-Alignment (BIA) and Modal Intra-Alignment (MIA). 
After the encoder outputs the Behavior representation $\overline{E}_{B}^u$ and $\overline{E}_{B}^i$, we use InfoNCE to maximize the mutual information between the user representation and item representation in behavior domain:
\begin{equation}
\mathcal{L}_{\text{BIA}} = \sum_{(u, i) \in \mathcal{B}} -\text{log}\frac{\exp(\overline{e}_{B}^u \cdot \overline{e}_{B}^i / \tau)}{\sum_{i' \in \mathcal{B}} \exp(\overline{e}_{B}^{u} \cdot \overline{e}_{B}^{i'} / \tau)} 
\end{equation} 

where $\tau$ is the temperature hyperparameter.
Similarly, we also utilize InfoNCE loss to align the Modal user representation and item representation in Modal Intra-Alignment. The Intra-Alignment Loss is the combination of Behavior Intra and Modal Intra Alignment:
\begin{equation}
\begin{aligned}
   \mathcal{L}_{\text{MIA}} &= \sum_{(u, i) \in \mathcal{B}} -\text{log}\frac{\exp(\overline{e}_{M}^u \cdot \overline{e}_{M}^i / \tau)}{\sum_{i' \in \mathcal{B}} \exp(\overline{e}_{M}^{u} \overline{e}_{M}^{i'} / \tau)} \\
   \mathcal{L}_{\text{Intra}} &= \mathcal{L}_{BIA} + \mathcal{L}_{MIA}
\end{aligned}
\end{equation}
\subsubsection{\textbf{Inter-Alignment}}
Until now, we get the Behavior representation $\overline{E}_{B}^u$, $\overline{E}_{B}^i$ and Modal representation $\overline{E}_{M}^u$, $\overline{E}_{M}^i$. However, representations of two lines do not share the same latent space. Therefore, in order to align behavior and modal information, we utilize the InfoNCE to align dual user/item representations obtained from Behavior Line and Modal Line:
\begin{equation}\label{Inter-Alignment Equation}
    \begin{aligned}
    \mathcal{L}_{\text{Inter}} &= \sum_{u \in \mathcal{B}}-\text{log}  \frac{\exp(\overline{e}_{B}^u \cdot  \overline{e}_{M}^u / \tau)}{\sum_{u' \in \mathcal{B}} \exp(\overline{e}_{B}^{u} \cdot \overline{e}_{M}^{u'} / \tau)} \\
     &+ \sum_{i \in \mathcal{B}}-\text{log}  \frac{\exp(\overline{e}_{B}^i \cdot \overline{e}_{M}^i / \tau)}{\sum_{i' \in \mathcal{B}} \exp(\overline{e}_{B}^{i} \cdot \overline{e}_{M}^{i'} / \tau)}
    \end{aligned}
\end{equation}
It is worth mentioning that in spite of the alignments between user and item representations from different lines, Behavior and Modal representations still retain their distinctiveness in terms of semantic information, which is further proven in Experiment section.  


\subsubsection{\textbf{General Optimazation}}
After we conduct Intra-Alignment and Inter-Alignment, we can get the general representations by just adding dual representations from Behavior (\ref{Behavior Line Equation}) and Modal lines (\ref{Modal Line Equation}):
\begin{equation}
\begin{aligned}
    E_{G}^u &= \overline{E}_{B}^u + \overline{E}_{M}^u \\
    E_{G}^i &= \overline{E}_{B}^i + \overline{E}_{M}^i 
\end{aligned}
\end{equation}
Finally, we adopt the Bayesian Personalized Ranking (BPR) Loss~\cite{rendle2012bpr} as general optimazation:
\begin{equation}
    \mathcal{L}_{\text{general}} = \sum_{(u,i,i') \in \mathcal{B}} -\log(\sigma(e_{G}^u\cdot e_{G}^i - e_{G}^u\cdot e_{G}^{i'}))
\end{equation} where each triple $(u,i,i')$ satisfies $A_{ui}=1$ and $A_{ui'}=0$, $\sigma(\cdot)$ is the sigmoid function. 
Our final loss is:
\begin{equation}
\begin{aligned}
    \mathcal{L}&= \mathcal{L}_{\text{general}} + \alpha \mathcal{L}_{\text{Intra}} + \beta \mathcal{L}_{\text{Inter}} +  \gamma \mathcal{L}_{\text{S3}} 
\end{aligned}
\end{equation}


\section{Experiment}\label{experiments}

We conduct comprehensive experiments to evaluate the performance and answer the following questions:
\begin{itemize}
    \item \textbf{RQ1}: How does DREAM perform compared with the existing multimodal recommendation methods?
    \item \textbf{RQ2}: What is the relationship between dual representations and how do they impact the model's performance?
    \item \textbf{RQ3}: How do the different modules (e.g. Modal-Specific Encoder, S3, BMA) influence the performance of the DREAM?
\end{itemize}

\begin{table}[ht]
\setlength{\abovecaptionskip}{-0cm}
\setlength{\belowcaptionskip}{-0.1cm}
\centering
    \caption{Statistics of the experimental datasets.}
    \label{Statistics of datasets}
\begin{tabular}{c|c|c|c|c}
\hline
Datasets    & \# Users & \# Items & \# Interactions & Sparsity \\ \hline
Baby        & 19, 445  & 7, 050   & 160, 792        & 99.88\%  \\
Sports      & 35, 598  & 18, 357  & 296, 337        & 99.95\%  \\
Clothing & 39, 387 & 23, 033   & 278, 677     & 99.97\%  \\ \hline
\end{tabular}
\end{table}

\begin{table*}[h]
\setlength{\abovecaptionskip}{-0cm}
\setlength{\belowcaptionskip}{-0cm}
    \centering
    \caption{Overall performance achieved by different recommendation methods in terms of Recall and NDCG. We mark the global best results on each dataset under each metric in boldface and the second best is underlined. }
    \label{Overall_performance}
    \resizebox{1.0\textwidth}{!}
{
\begin{tabular}{cc|cc|ccccccc|c}
\hline
\multicolumn{2}{c|}{}              & \multicolumn{2}{c|}{General Model} & \multicolumn{7}{c|}{Multi-modal  Model}                                               & Ours   \\ \hline
Dataset                   & Metric & BPR             & LightGCN         & VBPR   & DualGNN & SLMRec & BM3  & FREEDOM & LGMRec & MGCN     & DREAM    \\ \hline
\multirow{4}{*}{Baby}                  & R@10   & 0.0357          & 0.0479           & 0.0423 & 0.0448 & 0.0521 & 0.0564 &  0.0627 & \underline{0.0644} & 0.0620  & \textbf{0.0687} \\
                          & R@20   & 0.0575          & 0.0754           & 0.0663 & 0.0716 & 0.0772 & 0.0883 & 0.0992 & \underline{0.1002}&  0.0964 &\textbf{0.1040} \\
                          & N@10   & 0.0192          & 0.0257           & 0.0223 & 0.0240 & 0.0289 & 0.0301 & 0.0330  & \underline{0.0349}    &  0.0339 & \textbf{0.0360} \\
                          & N@20   & 0.0249          & 0.0328           & 0.0284 & 0.0309 & 0.0354 & 0.0383 & 0.0424  & \underline{0.0440}   &  0.0427 & \textbf{0.0459} \\
                          \hline
\multirow{4}{*}{Sports}   & R@10   & 0.0432          & 0.0569           & 0.0558 & 0.0568 & 0.0663  & 0.0656 & 0.0717 & 0.0720   &  \underline{0.0729} & \textbf{0.0776} \\
                          & R@20   & 0.0653          & 0.0864           & 0.0857 & 0.0859 & 0.0990 & 0.0980 & 0.1089  &   0.1068 &  \underline{0.1106} &  \textbf{0.1182} \\
                           & N@10   & 0.0241          & 0.0311           & 0.0307 &     0.0310 & 0.0365 & 0.0355 & 0.0385  & 0.0390    &  \underline{0.0397} & \textbf{0.0425} \\
                          & N@20   & 0.0298          & 0.0387          & 0.0383 & 0.0385& 0.0450 & 0.0438 & 0.0481  & 0.0480   &  \underline{0.0496} & \textbf{0.0529} \\ \hline
\multirow{4}{*}{Clothing} & R@10   & 0.0187          & 0.0340           & 0.0280 & 0.0454 & 0.0442 & 0.0437 & 0.0629 & 0.0555     &  \underline{0.0641} & \textbf{0.0683} \\
                          & R@20   & 0.0279          & 0.0526           & 0.0414 & 0.0683 & 0.0659  & 0.0648 & 0.0941  & 0.0828   &  \underline{0.0945} & \textbf{0.0991} \\
                          & N@10   & 0.0103          & 0.0188           & 0.0159 & 0.0242 & 0.0241 & 0.0247 & 0.0341  &  0.0302 &  \underline{0.0347} & \textbf{0.0367} \\
                          & N@20   & 0.0126          & 0.0236           & 0.0193 & 0.0299 & 0.0296  & 0.0302 & 0.0420  & 0.0371   &  \underline{0.0428} & \textbf{0.0448}\\ \hline
\end{tabular}
}
\end{table*}

\vspace{-0.5cm}
\subsection{Experimental Datasets} 
Following MMRec~\cite{MMrec,BM3}, we use the Amazon review dataset~\cite{amazon_review} for experimental evaluation. We choose three per-category datasets, i.e., Baby, Sports, and Clothing. The raw data of each dataset are pre-processed with a 5-core setting on both items and users, and their 5-core filtered results are presented in Table \ref{Statistics of datasets}. We use the pre-extracted 4,096-dimensional visual features and 384-dimensional text features, which have been published. 
\vspace{-0.3cm}
\subsection{Compared Methods} 
We compare DREAM with following recommendation methods. General CF models: BPR~\cite{rendle2012bpr}, LightGCN~\cite{he2020lightgcn}.  Multimodal recommendation model: VBPR~\cite{vbpr}, BM3~\cite{BM3}, DualGNN~\cite{dualgnn}, FREEDOM~\cite{MICRO}, LGMRec~\cite{guo2023lgmrec}, MGCN~\cite{MGCN}.
We use Recall@K(R@K) and NDCG@K(N@K) to evaluate the top-K recommendation performance of different recommendation methods.

\subsection{Effectiveness of DREAM (RQ1)}
The performances achieved by different models are summarized in Table \ref{Overall_performance}. DREAM significantly outperforms all baselines and achieves optimal results across different datasets. Specifically, DREAM improves the best baselines by 6.68\%, 6.45\%, 6.55\% in terms of Recall@10, 6.19\%, 7.05\%, 5.76\% in terms of NDCG@10 on Baby, Sports and Clothing. We attribute these substantial performance improvements to three key factors: i) The Modal-specific Encoders utilizes modal information more sufficiently through filter gate and relation graphs; ii) The Similarity Supervised Signal mitigate the problem of Modal Information Forgetting; iii) The Behavior-Modal Alignment (BMA) module brings the dual representations closer together, thereby simplifying the information fusion process.

\subsection{In-Depth Analysis on Dual Representations (RQ2)}\label{analysis on dual representation}
\subsubsection{\textbf{The Recommendation Performace of Dual Representation}}
\begin{figure}[ht]
  \setlength{\abovecaptionskip}{-0cm}
  \setlength{\belowcaptionskip}{-0.1cm}
  \includegraphics[width=0.5\textwidth]{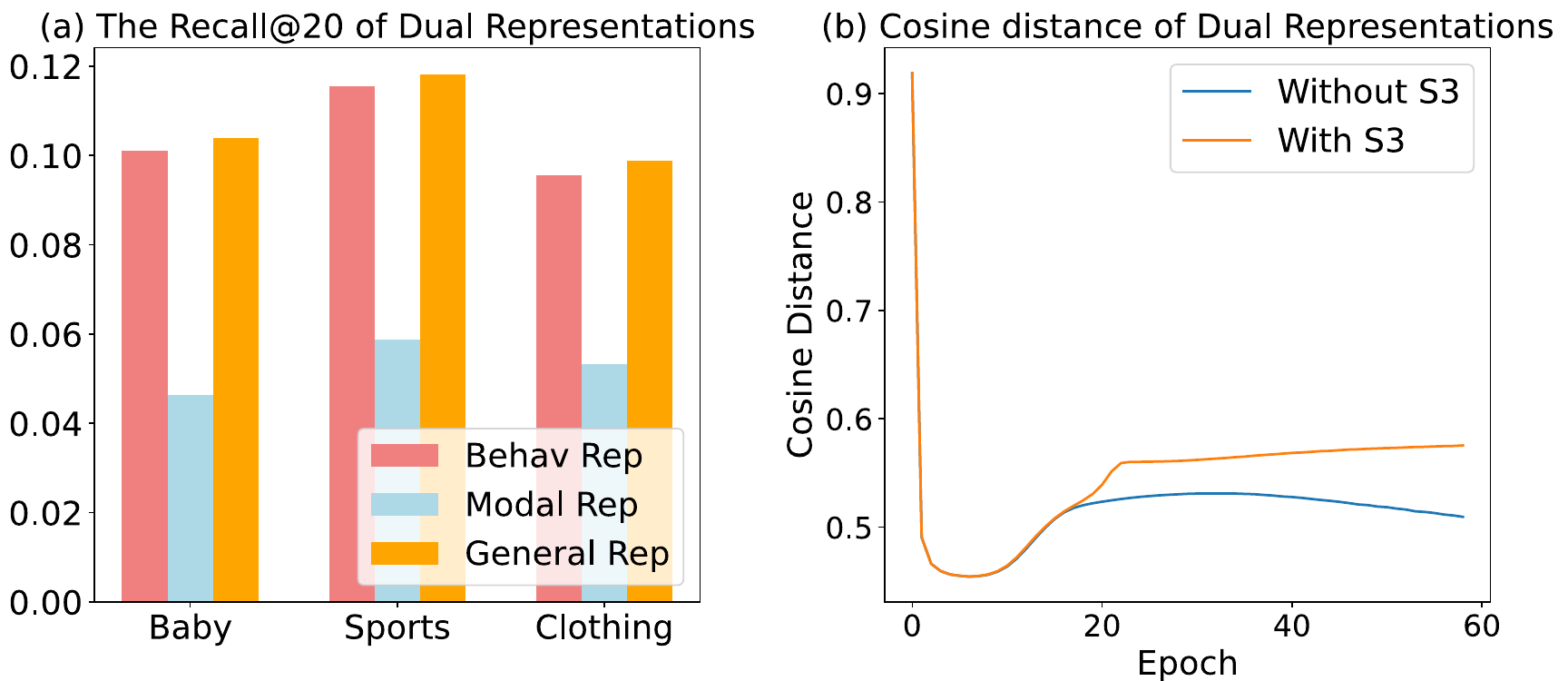}
  \caption{(a) The Recommendation performance and (b) Cosine distance of Dual Representation.}
  \label{dual_representation performance}
\end{figure}

To investigate the individual impacts of the dual representations, we attempted to directly use Behavior representation and Modal representation for recommendation evaluation, with the result shown in Figure \ref{dual_representation performance}(a). We can find that The performance of Behavior representation is generally better than Modal representation. Behavior information plays a central role in recommendation systems. In contrast, multimodal information typically serves a supplementary role in enhancing recommendation effectiveness.

\subsubsection{\textbf{The Cosine Distance between Dual Representations}} We present cosine distance between Behavior representation and Modal representation during learning in Figure \ref{dual_representation performance}(b). 
In early training epochs(0-10), the cosine distance between behavior and modal representations decreases rapidly. This is due to the significant initial disparity in the distributions of the dual representations, which leads to a substantial alignment loss, which dominates the parameter updates, facilitating a quick convergence of the representations towards each other. However, this also impedes their ability to independently capture distinct semantic meanings.
As training progresses (after Epoch 10), the behavior and modal representations become more semantically independent. The cosine distance gradually stabilizes, indicating model convergence. We also examine the impact of S3. By incorporating S3, the modal representation is made to capture the similarity signals in the original multimodal features, further preserving semantics. This results in a larger cosine distance and significantly improved model performance.


\subsubsection{\textbf{The Visualization of Dual Representation Distribution}} 
\begin{figure}[ht]
  \centering
  \begin{subfigure}[b]{0.22\textwidth}
    \centering
    \includegraphics[width=\textwidth]{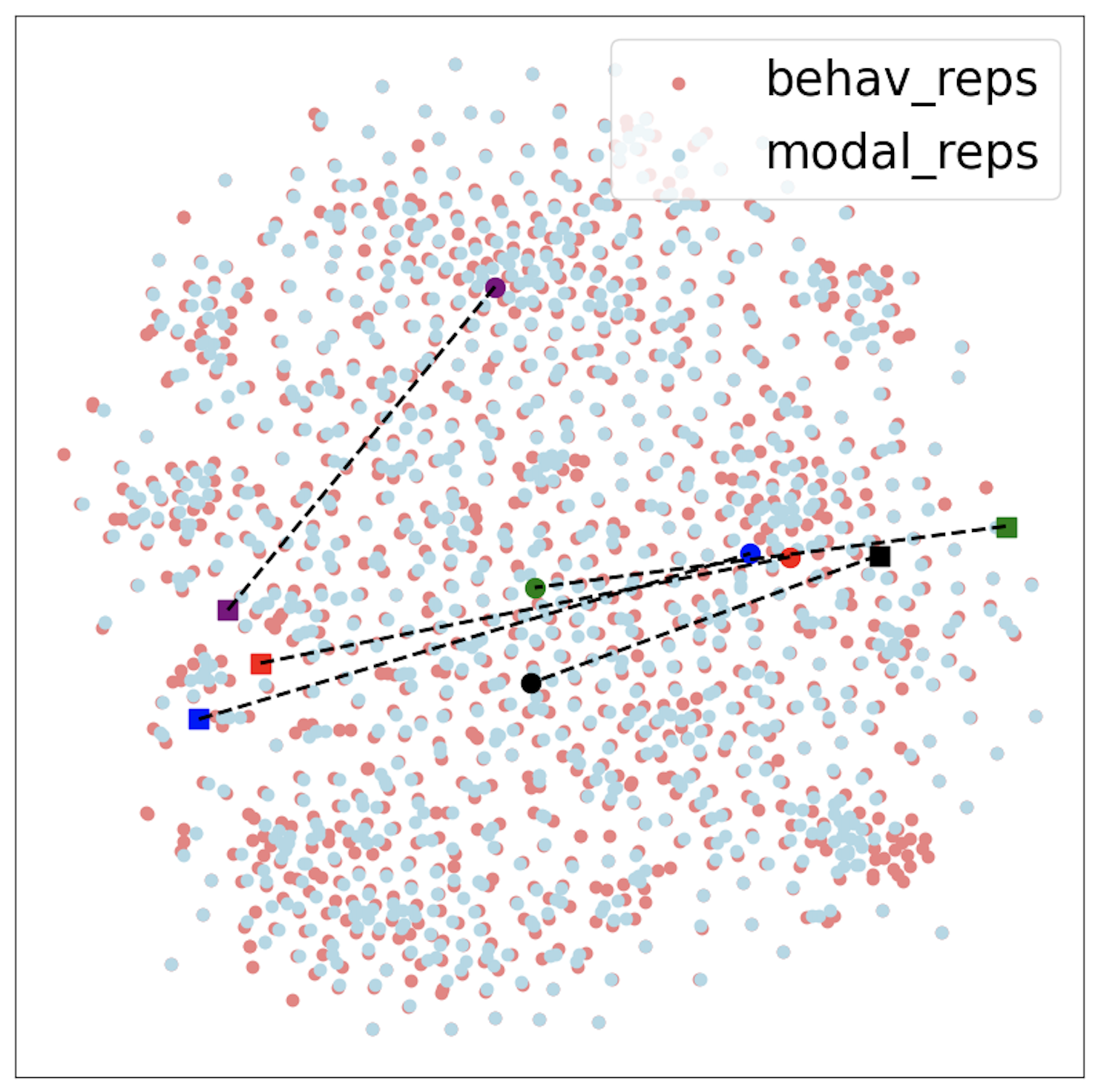}
    \caption{Sports}
    \label{distribution_behavior}
  \end{subfigure}
  \hfill
  \begin{subfigure}[b]{0.22\textwidth}
    \centering
    \includegraphics[width=\textwidth]{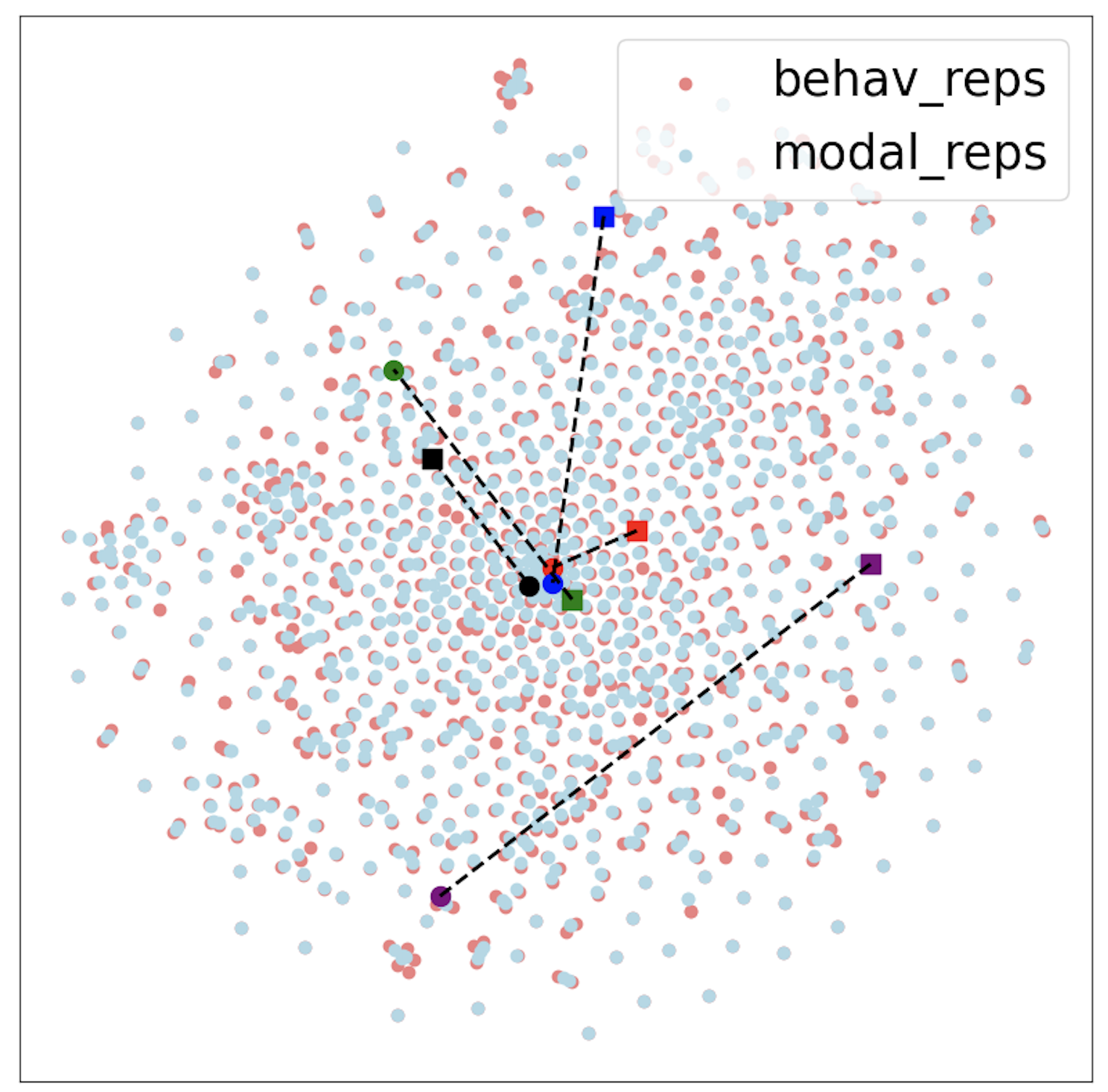}
    \caption{Clothing}
    \label{distribution_modal}
  \end{subfigure}
  \caption{The 2D distributions of learned Behavior representation and Modal representation in DREAM. The circles and squares of the same color represent the dual representations corresponding to the same item or user. This illustrates that DREAM effectively aligns the information across the dual domains while also preserving the independent semantics of the dual representations.}
  \label{distribution}
\end{figure}

We randomly selected 1000 instances (items and users) from the Sports and Clothing dataset and visualized the two-dimensional distributions of the Behavior representation and Modal representation generated by T-SNE~\cite{van2008t-sne} in Figure \ref{distribution}, which shows that the dual representations have converged to similar overall distributions. 
The dashed line connects the behavior representation (square) with the modality representation (circle) of the same item, indicating that there is still a significant distance between the dual representations of the same item. 
These observations demonstrate the capacity of DREAM to \textbf{align distinct features in dual domains while also preserving the independent semantics of each representation}.


\subsection{Ablation Study (RQ3)}
\begin{table}[]
\setlength{\abovecaptionskip}{-0cm}
\setlength{\belowcaptionskip}{-0.05cm}
\caption{The Recall@20 and NDCG@20 of Ablation Study}
\label{RQ3}
\resizebox{0.5\textwidth}{!}{
\begin{tabular}{c|cc|cc|cc}
\hline \hline
                           & \multicolumn{2}{c|}{Baby} & \multicolumn{2}{c|}{Sports} & \multicolumn{2}{c}{Clothing} \\ \hline
                        Metric   & R@20         & N@20        & R@20          & N@20         & R@20           & N@20          \\ \hline \hline
 DREAM                       & \textbf{0.1040}       & \textbf{0.0459}      &\textbf{ 0.1182    }    & \textbf{0.0529  }     & \textbf{0.0991 }        & \textbf{0.0448}        \\ \hline \hline
 w/o Filter Gate & 0.0879 & 0.0374 & 0.0987 & 0.0437 & 0.0799 & 0.0362 \\ 
 w/o Relation Graphs & 0.0812 & 0.0345 & 0.0967 & 0.0414 & 0.0803 & 0.0305 \\
 w/o Text Encoder & 0.0884 & 0.0385 & 0.0976 & 0.0444 & 0.0711 &  0.0321 \\
 w/o Image Encoder & 0.1008 & 0.0441 & 0.1124 & 0.0506 & 0.0817 & 0.0367 \\
 w/o Text\&Image Encoders & 0.0854 & 0.0375 & 0.0963 & 0.0408 & 0.0675 & 0.0302 \\
 \hline \hline
 w/o S3& 0.0976 & 0.0434 & 0.1103 &  0.0491 & 0.0987 & 0.0443\\ \hline \hline
 w/o Inter-Alignment       & 0.0971       & 0.0433      & 0.1114       & 0.0497      & 0.0912         & 0.0406        \\
 w/o Intra-Aligment       & 0.0985       & 0.0442      & 0.1125        & 0.0507       & 0.0951         & 0.0431        \\
 w/o Alignment             & 0.0899      & 0.0405      & 0.0965       & 0.0427      & 0.0806         & 0.0360        \\ \hline \hline
\end{tabular} 
}
\end{table}

We design ablation experiments to investigate the influence of different modules on model performance. From the results in Table \ref{RQ3}, we can observe: (1) The Modal-specific encoder including filter gate and relationship graph is important for Modal representation learning. Because removing either of them can cause performance drops; (2) Multimodal information is crucial, as removing either the text or image encoder leads to significant performance degradation. The performance drop is more pronounced when the text encoder is removed, suggesting that the textual modality provides greater benefits for the recommendation task; (3) w/o S3 reflects the contribution of S3 to the model performance. Additionally, Fig. \ref{fig:information_shift} shows S3 can mitigate the problem of Modal Information Forgetting. Fig. \ref{dual_representation performance}(b) further illustrates the effectiveness of S3 in preserving the semantic information of modal representation; (4) The Behavior-Modal Alignment (BMA) module includes Intra-Alignment and Inter-Alignment, which both can improve performance. w/o Inter-Alignment gets more performance drop, which indicates Inter-Alignment (aligning representations from dual domains) is more important for information fusion.
Furthermore, we integrate the BMA into other recommedation models, with the detailed results shown in Fig. \ref{fig:bma_introduction} and Table \ref{align_improvement}.

\section{Conclusion}
In this study, we propose DREAM, which uses two independent lines of representation learning to calculate behavior and modal representations. Especially, we utilize Modal-specific Encoder including filter gates and relation graphs for fine-grained modal representation learning of users and items. We identify the problem of Modal Information Forgetting, and introduce the Similarity Supervised Signal, which encourages the modal representation to maintain the similarity information embodied in original features. We design the Behavior-Modal Alignment (BMA) for misalignment problem, and integrate it into existing recommendation models, leading to consistent performance improvement.\footnote{Due to the page limit, the detailed introduction of baseline models, implementation details, the parameter sensitivity study, the computation complexity analysis and convergency curve are all provided in the Supplementary Material.}
\begin{table}[]
\caption{ The performance improvement of integrating the BMA into other models, while keeping their hyper-parameters and model structure unchanged.}\label{align_improvement}
\resizebox{0.5\textwidth}{!}{
\begin{tabular}{c|cc|cc|cc}
\hline
       & \multicolumn{2}{c|}{Baby}                                & \multicolumn{2}{c|}{Sports}       & \multicolumn{2}{c}{Clothing}      \\ \hline
Models & R@20                                   & N@20            & R@20            & N@20            & R@20            & N@20            \\ \hline
VBPR   &  0.0663          & 0.0284          & 0.0857          & 0.0383          & 0.0414          & 0.0193          \\
+ BMA  & \textbf{0.0682} & \textbf{0.0292} & \textbf{0.0874} & \textbf{0.0388} & \textbf{0.0638} & \textbf{0.0282} \\ \hline
BM3    & 0.0883                                 & 0.0383          & 0.0980          & 0.0438          & 0.0648          & 0.0302          \\
+ BMA  & \textbf{0.0907}                        & \textbf{0.0387} & \textbf{0.1078} & \textbf{0.0494} & \textbf{0.0909} & \textbf{0.0414} \\ \hline
MGCN   & 0.0964                                 & 0.0427          & 0.1106          & 0.0496          & 0.0945 & 0.0428 \\
+ BMA  & \textbf{0.1013}                        & \textbf{0.0435} & \textbf{0.1161} & \textbf{0.0513} & \textbf{0.0949}        & \textbf{0.0433    }   \\ \hline
\end{tabular}
}
\end{table}

\bigskip

\bibliography{aaai25}

\section{Appendix}
\section{Baseline Models}
\subsection{General Models}
\begin{itemize}
    \item BPR~\cite{rendle2012bpr}: This method utilize the historical user-item interactions to model user and item representations as the latent factor. This model introduces the mainly used loss function BPR loss and predicts user's preference based on the similarity between the representations.
    \item LightGCN~\cite{he2020lightgcn}: This is the most popular GCN-based colaborative filtering mothod, which simplifies the design of GCN to make it more appropriate for the recommendation. 
\end{itemize}
\subsection{Multimodal Models}
\begin{itemize}
    \item VBPR~\cite{vbpr}: This model integrates the visual features and ID embeddings of each item as its representation. To be fair, we concatenate both vision and text features as the multimodal feature when we learn the representations. The BPR loss is used here to learn the user preference. 
    \item DualGNN~\cite{dualgnn}: This model introduces a new user-user co-occurrence graph and utilize it to fuse the user representation from its neighbors 
    in the correlation graph. 
    \item SLMRec~\cite{SLMRec}: This model incorporates self-supervised learning into multimodal recommendation, where design three data augmentations to uncover the multimodal patterns in data for contrastive learning. 
    \item BM3~\cite{BM3}: This method simplifies the self-supervised approach. It removes the requirement of randomly sampled negative examples and directly perturbs the representation through a dropout mechanism. 
    \item FREEDOM~\cite{Freedom}: The model utilize the item-item graph for each modality established the same as LATTICE but freezes the graphs before training, and introduces the degree-sensitive edge pruning techniques to denoise the user-item interaction graph. 
    \item LGMRec~\cite{guo2023lgmrec}: The model introduces the local and global graph learning for jointly modeling local and global user interests. 
    \item MGCN~\cite{MGCN}: The model propose a novel Multi-View Graph Convolutinal Network, equiping three specically designed modules for the problem of modality noise and incomplete user perference modeling. 
\end{itemize}

\section{Implementation Details}
We implement DREAM and all the baselines with MMRec~\cite{MMrec} framework. To ensure fair compareness, we fix the embedding size of both users and items to 64, initialize the embedding parameters with the Xavier method\cite{Xavier}, and use Adam~\cite{kingma2014adam} as the optimizer with a learning rate of 0.001. Specifically, the GCN layer in Behavior Line is fixed as 2, the layer in Modal Line is fixed as 1, the regularization coefficient $\lambda$ is set as $10^{-4}$, the batch size is set to $B=2048$ and the temperature $\tau =0.2$.
Further, the weight of Intra-Alignment $\alpha$ is searched from $\{0.01, 0.03\}$, the weight of Inter-Alignment $\beta$ is searched from $\{0.01, 0.03\}$,  $\gamma$ is searched from $\{0.1, 0.5\}$ and $\alpha_v$ is searched from $\{0.1, 0.3, 0.5\}$. 
The early stopping and total epochs are fixed at 20 and 1000, respectively. Following ~\cite{lattice}, we use Recall@20 on the validation data as the training-stopping indicator.
\section{Sensitivity Analysis}
\begin{figure}[ht]
  \setlength{\abovecaptionskip}{-0cm}
  \setlength{\belowcaptionskip}{-0.1cm}
  \centering
  \includegraphics[width=0.48\textwidth]{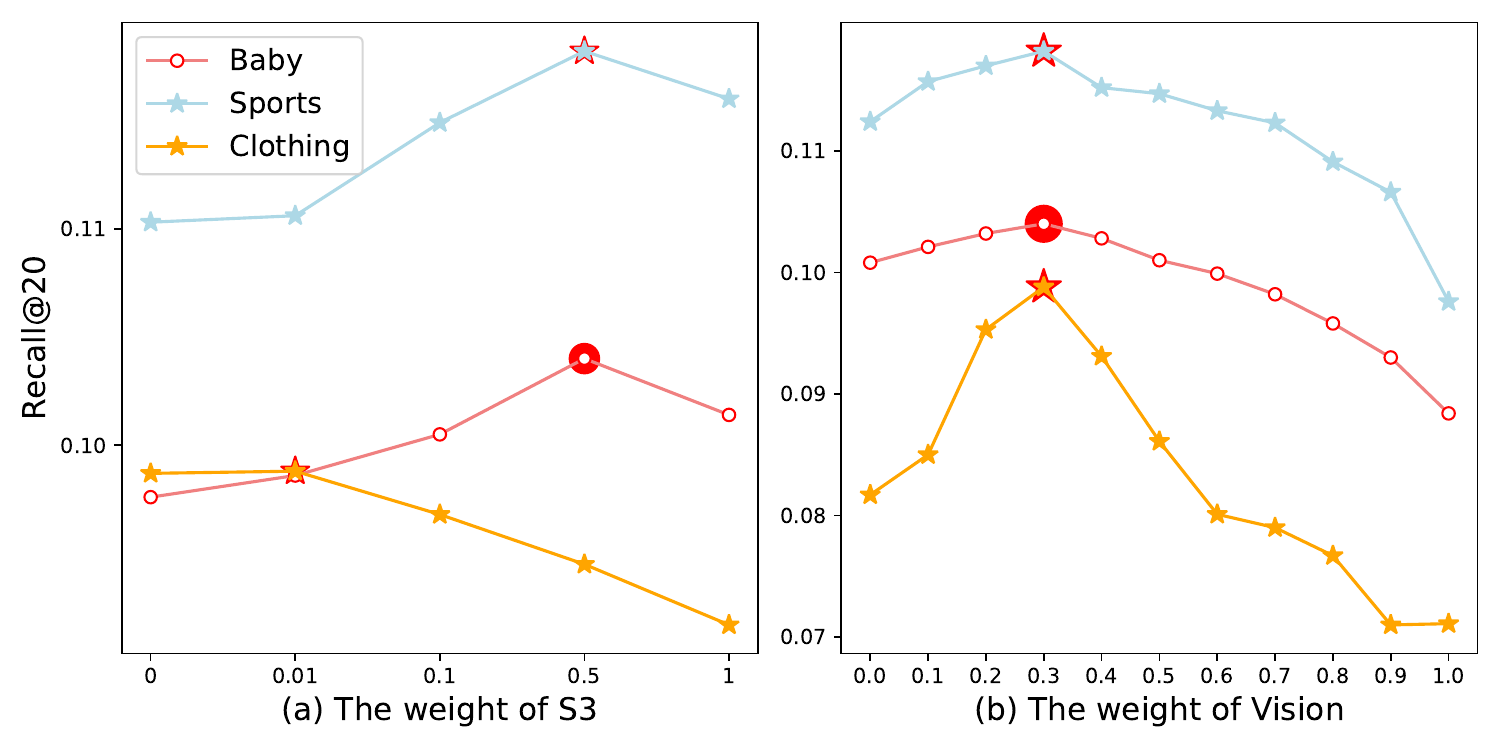}
  \caption{The Recall@20 results of different S3 weight and vision weight.}
  \label{sensetive performance}
\end{figure}

\subsubsection{\textbf{The weight of S3}}
Based on the analysis of Figure \ref{sensetive performance} (a), we observed that the optimal performance of the S3 weight varies across different datasets. Interestingly, the model's performance initially improves and then deteriorates as the weight of S3 increases. This suggests that the S3 signal is indeed beneficial for enhancing the model's performance. However, excessively high weights assigned to S3 result in an overemphasis on capturing similarity information within modality representation. Consequently, the model's ability to capture behavior information might be compromised, leading to a decline in overall performance.

\subsubsection{\textbf{The weight of modal weight}}
From Figure ~\ref{sensetive performance} (b), we find that both textual and visual features can improve performance. By combining both the textual and visual features, DREAM achieves best recommendation performance. In all three datasets, best vision weight is $0.3$. It means text modality is more important than  vision modality, which is consistent with BM3~\cite{BM3}.

\begin{figure}[ht]
  \setlength{\abovecaptionskip}{-0cm}
  \setlength{\belowcaptionskip}{-0.1cm}
  \centering
  \includegraphics[width=0.5\textwidth]{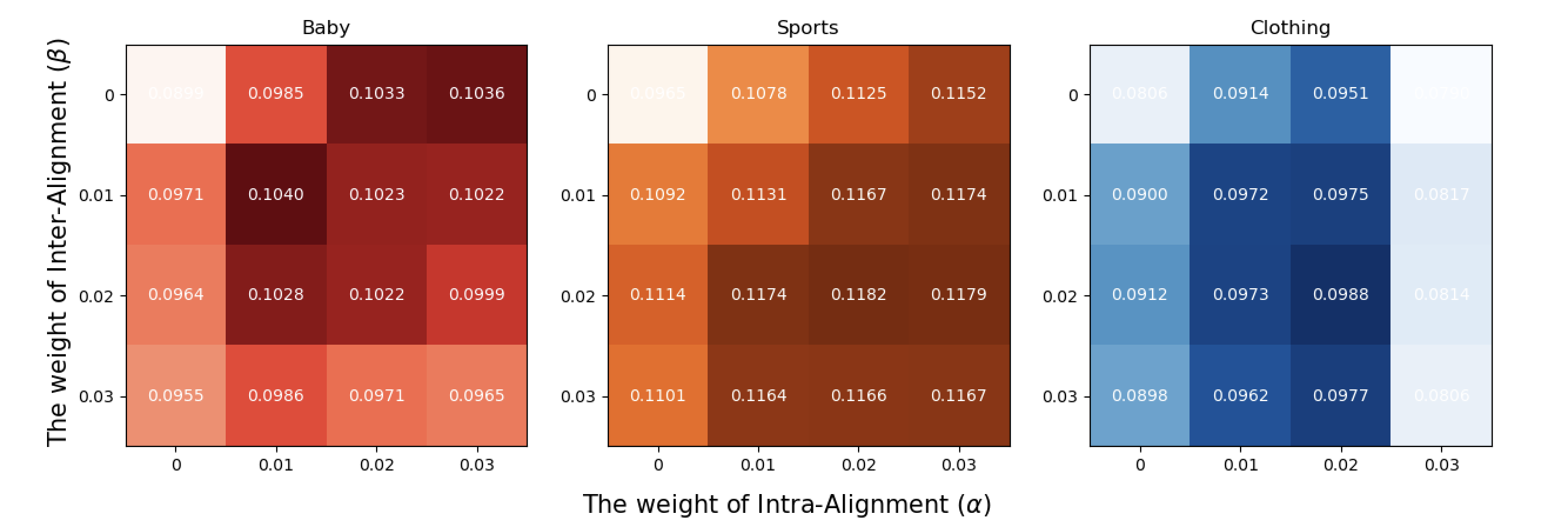}
  \caption{The Recall@20 results of different Intra-alignment and Inter-alignment weight.}
  \label{hotmap figure}
\end{figure}

\subsubsection{\textbf{The weight of Intra-alignment and Inter-alignment weight}}
In our study, we propose the Behavior-Modal Alignment module for achieving both Intra-Alignment and Inter-Alignment.
In this section, we examine the impact of weights, denoted as $\alpha$ for Intra-Alignment and $\beta$ for Inter-Alignment. The results obtained from three datasets are presented in Figure ~\ref{hotmap figure}. 

\section{The Behavior-Modal Alignment based on other models}
\begin{figure}[ht]
  \centering
  \includegraphics[width=0.48\textwidth]{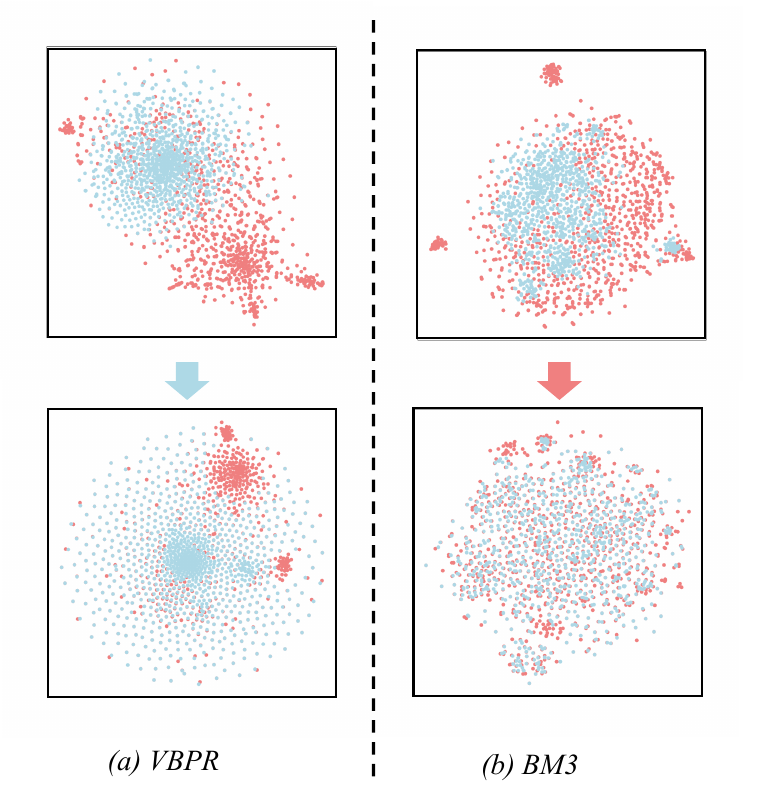}
  \caption{The representation distribution transformation after introducing the BMA into VBPR~\cite{vbpr} and BM3~\cite{BM3}}
  \label{fig:distribution_transform}
\end{figure}

In this section, we utilize T-SNE~\cite{van2008t-sne} to visualize the distribution transformation after integrating the Bahabvior-Modal Alignment module into previous work (e.g. VBPR~\cite{vbpr}, BM3~\cite{BM3}), with the results shown in Figure \ref{fig:distribution_transform}. From Figure \ref{fig:distribution_transform}, we could infer that the Behavior-Modal Alignment (BMA) effectively align the behavior-related representations with the modal-related representations. 

\section{Computation Effectiveness}
\subsection{Computation Complexity}
The computational cost of DREAM mainly focus on dual line representation learning, the feature transform and loss computation. 
DREAM pre-buids the Modal-specific U-U and I-U graphs before training and freezes them during training so the computional cost is LightGCN on behavior graph and modal graphs. The cost of LightGCN on these graphs is respectively $\mathcal{O}(2L|\mathcal{E}d/B|)$, $\mathcal{O}(kd|\mathcal{U}|)$ and $\mathcal{O}(kd|\mathcal{I}|)$, where $L$ is the number of LightGCN layers and $B$ is the training batch size. 
The overall loss cost are $\mathcal{O}((4 + 2|\mathcal{M}|)dB + |\mathcal{M}|d_mB )$ including Intra-Alignment, Inter-Alignment and Similarity Supervised Signal. All in all, based on previous work~\cite{BM3,Freedom}, we just introduce additional linear computational overhead. However, this modification led to the model achieving best performance across all datasets. We summarize the computational complexity on multi-modal graph-based methods in Table \ref{Computational Complexity}.

\begin{table}[htp]
    \centering
    \caption{Comparison of computational complexity. We set $X=2L|\mathcal{E}d/B|$.}
    \vspace{-2mm}
    \resizebox{0.5\textwidth}{!}{
    \begin{tabular}{c|c|c|c} 
        \hline
        \textbf{Models}  & \textbf{BM3} &\textbf{ FREEDOM} & \textbf{DREAM} \\ \hline
        \textbf{Graph Convolution} &  $\mathcal{O}(X)$ & $\mathcal{O}(X + kd|\mathcal{I}|)$ & $\mathcal{O}(X + kd|\mathcal{I} + \mathcal{U}| )$ \\ 
        \textbf{Feature Transform}  & $\mathcal{O}(\sum_{m \in \mathcal{M}} |\mathcal{I}|d_md) $ & $\mathcal{O}(\sum_{m \in \mathcal{M}} |\mathcal{I}|d_md) $ & $\mathcal{O}(\sum_{m \in \mathcal{M}} |\mathcal{I}|d_md)$  \\ 
       \textbf{ Losses} & $\mathcal{O}((2 + 2|\mathcal{M}|)dB)$ & $\mathcal{O}((2 + 2|\mathcal{M}|)dB)$ & $\mathcal{O}((4 + 2|\mathcal{M}|)dB) $\\ \hline
    \end{tabular}
    }
    \label{Computational Complexity}
\end{table}

\subsection{Convergency Curve}
\begin{figure}[ht]
  \setlength{\abovecaptionskip}{-0cm}
  \setlength{\belowcaptionskip}{-0.1cm}
  \centering
  \includegraphics[width=0.5\textwidth]{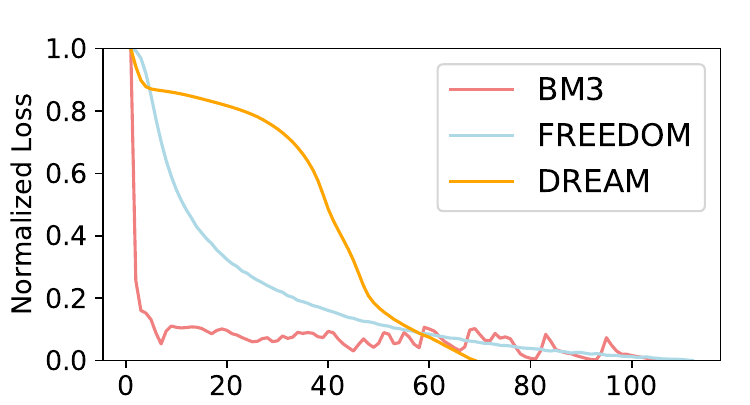}
  \caption{Loss convergency curves of three models in Baby dataset. (X-axis means Epoch)}
  \label{fig:convergency curve}
\end{figure}

The loss convergency curves about three models in Baby dataset(e.g. BM3, FREEDOM, DREAM) are shown in Figure \ref{fig:convergency curve}. DREAM converges in just 69 epochs, significantly faster than the 106 and 112 epochs required for BM3 and FREEDOM, respectively.

\end{document}